\documentclass[conference]{IEEEtran}
\usepackage{cite}
\usepackage{amsmath,amssymb,amsfonts}
\usepackage{algorithmic}
\usepackage{graphicx}
\usepackage{textcomp}
\usepackage{multirow}
\usepackage[table,xcdraw]{xcolor}
\def\BibTeX{{\rm B\kern-.05em{\sc i\kern-.025em b}\kern-.08em
    T\kern-.1667em\lower.7ex\hbox{E}\kern-.125emX}}
\begin{document}

\title{Accuracy, Fairness, and Interpretability of Machine Learning Criminal Recidivism Models}

\author{
    \IEEEauthorblockN{Eric Ingram, Furkan Gursoy, Ioannis A. Kakadiaris}
    \IEEEauthorblockA{\textit{Computational Biomedicine Lab} \\
\textit{Dept. of Computer Science} \\
\textit{University of Houston}\\
Houston, TX, USA
    \\ emingram23@amherst.edu, \{fgursoy, ioannisk\}@uh.edu}
}

\maketitle
\thispagestyle{plain}
\pagestyle{plain}

\begin{abstract}
Criminal recidivism models are tools that have gained widespread adoption by parole boards across the United States to assist with parole decisions. These models take in large amounts of data about an individual and then predict whether an individual would commit a crime if released on parole. Although such models are not the only or primary factor in making the final parole decision, questions have been raised about their accuracy, fairness, and interpretability. In this paper, various machine learning-based criminal recidivism models are created based on a real-world parole decision dataset from the state of Georgia in the United States. The recidivism models are comparatively evaluated for their accuracy, fairness, and interpretability. It is found that there are noted differences and trade-offs between accuracy, fairness, and being inherently interpretable. Therefore, choosing the best model depends on the desired balance between accuracy, fairness, and interpretability, as no model is perfect or consistently the best across different criteria.
\end{abstract}


\section{Introduction}
When considering individuals for parole, many parole boards are now using machine learning (ML) models as a factor in their decision. Over time, these models have seen an increase in adoption. An early example is the state of Pennsylvania in 2010, where the state created a model that would predict criminal recidivism (whether or not an individual would commit a crime again while on parole) \cite{b1}. This model was not the sole deciding factor in whether an individual would be granted parole but was seen as another piece of evidence by the parole board that could be used in their decision-making. Arnold PSA \cite{b1.2} and COMPAS \cite{b1.3} are two popular examples of decision-aiding systems employed in different criminal justice agencies across the United States.

As expected with any ML model used for public policy, questions arise about accuracy, fairness, and interpretability in criminal recidivism models. Is the predictive performance acceptable? Is the model fair across protected groups and statuses (e.g., gender, race)? Can people understand \emph{why} a decision was made? Moreover, there needs to be some measure of trust in the model. The ``black-box'' approach---data plugged in and a prediction made with no explanation of what happened in between--- is insufficient for policymakers and the general public to understand and trust the decisions made by the model. 

There are also ethical issues involved that could be better addressed with improved fairness and interpretability in recidivism models. For example, it would be unethical to have a model decide that someone should be granted parole while someone else should not if all things are equal other than gender or race. Individuals must feel confident that the model is ethical across protected statuses.
Learning fair models is often difficult due to the imbalance of protected classes going through the parole board or historical biases.
Moreover, if the model decides that someone should be granted parole and commit a serious crime, is the model to blame? Is the parole board? It is easier to explain why someone was granted parole if the parole board can explain why a model made the decision it did.

To help tackle the fairness problem, the National Institute of Justice (NIJ), a United States government agency, created a challenge. The challenge's goal was for each team to submit a model trained on provided training data that aimed to be fair and accurate, hopefully helping to advance scientific knowledge in creating fair and accurate criminal recidivism models. The NIJ would then evaluate the model based on a private training set and select the winning teams for accuracy and fairness \cite{b2}. However, the challenge itself did not place emphasis on interpretability. It was possible that the winning models did not make predictions based on factors one would typically expect \cite{b3}. Some groups did use interpretability methods to analyze their models, but interpretability was not the primary objective and was not explored in depth. One winning team admitted to having ``gamed'' the fairness metric to win one of the fairness categories, ignoring the model's interpretability \cite{b4}. 

This paper presents several ML models that are trained on the NIJ data and then are subsequently analyzed for accuracy, fairness, and interpretability. The models are then compared and contrasted to examine the trade-offs between accuracy, fairness, and interpretability.

\section{Prior Work}

As part of the challenge, many ML-based recidivism prediction models were created \cite{b3, b4, b5} with the same data that this work utilizes. However, the models developed in this paper target an overall recidivism prediction, while the challenge models target a recidivism prediction for a particular year of the data \cite{b4}.

Berk \cite{b1} studied a retrospective analysis of the impact of criminal recidivism models in Pennsylvania. The analysis made no definite conclusions because the criminal recidivism models are not the sole factor in making the parole decisions. However, based on the data the paper analyzed, it is stated that there is no evidence to indicate that the use of the model harmed public safety overall.  

Wang et al. \cite{b6} developed interpretable and black-box criminal recidivism models on two different recidivism datasets. They then compared the fairness and interpretability of the models created against two models that are currently used in the justice system: Arnold PSA \cite{b1.2} and COMPAS \cite{b1.3}. They found that interpretable models can perform similarly to non-interpretable models and the two currently used models. 

\section{Methodology}

\subsection{Data}

The data provided by the NIJ contains information about individuals from the state of Georgia who were granted parole from 2013--2015 \cite{b2}. A wide variety of information is provided about these individuals. Criminal records, drug testing, employment, prior parole, and demographic data are some groups of available variables. The target variable of interest in this work is \emph{Recidivism\_Within\_3years}. This binary variable indicates whether an individual committed a crime again after being released on parole during the three-year window being studied. 
It is important to note that this variable is slightly unbalanced, with 58\% of the individuals committing a crime in the three years following the granted parole. Three other target variables are provided for the challenge prizes, where participating teams are asked to predict if someone would commit a crime each year. These yearly target variables are not considered in this paper. Finally, the NIJ provided an approximately 70/30 training/test split. 

In this work, the data was preprocessed to be usable across various ML models. Missing data was a prevalent issue in the dataset, with 45\% of rows having at least one missing value. Missing data were imputed using various methods, both naive and informed. Naive imputation strategies (i.e., imputation using descriptive statistics such as mean, median, and mode) were employed when there was a low correlation between variables and when other variables could not accurately predict the variable with the missing values. For example, the majority of missing values were from \emph{Avg\_Days\_per\_DrugTest}: how often an individual was tested for drugs. Despite other drug-related features, there was little correlation to other features, and a fitted linear regression model was not statistically significant. Therefore, missing values of \emph{Avg\_Days\_per\_DrugTest} were imputed using the median value. For informed imputation, one of the following strategies is used: (i) assumption and (ii) ML models. Some missing values could be assumed. For example, if \emph{Jobs\_Per\_Year} was 0, logically \emph{Percent\_Days\_Employed} should be 0 and vice versa. Some missing values may be predicted from other features using an ML model. For example, \emph{Supervision\_Level\_First} was correlated with \emph{Supervision\_Risk\_Score\_First} ($\rho = 0.53$), and a K-Nearest Neighbors model could predict \emph{Supervision\_Level\_First} from \emph{Supervision\_Risk\_Score\_First} with $0.70$ accuracy. Therefore, missing values of \emph{Supervision\_Level\_First} were imputed using the predictions of the K-Nearest Neighbors model. 

Ordinal variables were encoded numerically where the lowest ranked category is assigned 0, the second lowest is assigned 1, and so on. Boolean variables were encoded using 0 for False and 1 for True. Non-binary categorical variables were converted to multiple binary variables by using \textit{k-1} one-hot encoding.

Further, some features were removed because they would have no use in the analysis: ID and unused target variables. Before preprocessing, the dataset contained 25,835 individuals (18,028 in the training set, 7,807 in the test set) and 54 variables. After preprocessing, the dataset contains 62 variables used as predictors and a single binary target variable indicating recidivism status within the three years following the parole decision.

\subsection{Models}

The objective is to create a wide variety of ML models, both inherently interpretable and not inherently interpretable, to allow for comparison. The two inherently interpretable models implemented are a decision tree \cite{b7} and a logistic regression model with L1 regularization (LASSO) \cite{b8}. The implemented models that are not inherently interpretable are an adaptive boosting classifier \cite{b9}, a gradient boosting classifier \cite{b10}, an XGBoost classifier \cite{b11}, a random forest classifier \cite{b12}, a support vector machine (SVM) \cite{b13}, and a multilayer perceptron neural network \cite{b14}. The best hyperparameter configuration for each model was found using a grid search over the parameter space with 5-fold cross-validation on the training set. Then, each model with its best hyperparameter values was trained on the entire training set. Sequentially, the model predictions were obtained for the test set.

\subsection{Fairness}

Multiple tools can be used to assess the fairness, one of the most popular being the Aequitas Bias and Fairness Audit Toolkit \cite{b15}, an open-source software developed and hosted by the University of Chicago. The Aequitas audit was conducted for all models using gender and race information, models' test set predictions, and the corresponding ground-truth values. All bias tests provided by Aequitas are included in the analysis. The tests are:
\begin{enumerate}
  \item \emph{Predicted Positive Rate Disparity (PPRD)}, whether the numbers of positive predictions are on par across groups.
  
  \item \emph{Predicted Positive Group Rate Disparity (PPGRD)}, whether the rates of positive predictions are on par across groups.

  \item \emph{False Discovery Rate Disparity (FDRD)}, whether the ratios of false positives to predicted positives are on par across groups.
  
  \item \emph{False Positive Rate Disparity (FPRD)}, whether the ratios of false positives to actual negatives are on par across groups.
  
  \item \emph{False Omission Rate Disparity (FORD)}, whether the ratios of false negatives to predicted negatives are on par across groups.
  
  \item \emph{False Negative Rate Disparity (FNRD)}, whether the ratios of false negatives to actual positives are on par across groups.
\end{enumerate}

\subsection{Interpretability}

As indicated before, a mix of inherently and not inherently interpretable models were implemented in this work. Being inherently interpretable means it is feasible to see precisely why a decision was made by just looking at the model itself \cite{b16}. Decision trees and general linear models are inherently interpretable.

Decision trees make their prediction by creating decision boundaries based on what minimizes the degree of uncertainty of the model. Therefore, the decisions made at the nodes of a tree can be examined to see how the nodes partition the dataset until the tree reaches a terminal node. Thus, it is trivial to track how an individual prediction is made and to see why a prediction was made (and what would change it). The typical explanation is a visualization of a tree with the corresponding decision rule indicated on each node. Alternatively, feature importance methods such as Gini importance \cite{b7} can be used to understand the model's decision-making. Gini importance measures the average gain of purity by splits of a given variable. The larger the Gini importance value for a variable, the more its contributions to the model.

LASSO, the other inherently interpretable model employed in this paper, is a regularized general linear model. The optimization procedure learns coefficients for all features. Each feature coefficient indicates the extent of influence of the one unit change in that feature on model predictions. Hence, such coefficients can be read to understand how a LASSO model arrives at its decisions.

Not inherently interpretable models are too complicated to reasonably interpret by inspection or impossible to interpret in the same way as inherently interpretable models. Therefore, additional methods are needed to interpret these models, which are more complicated than just observing the models' coefficients or Gini feature importances. The methods that are used in this paper are global surrogate models \cite{b16},  permutation feature importance \cite{b18}, Shapley Additive Explanations (SHAP) \cite{b19}, and Accumulated Local Effects (ALE) plots \cite{b20}. 

A global surrogate model \cite{b16} is an interpretability method that attempts to explain the predictions of a black-box model by using an inherently interpretable model as a surrogate model. However, the surrogate model is trained to learn the black-box model's predictions rather than the ground-truth values. Then, the surrogate model, being inherently interpretable, can explain the original black-box model's predictions in general. However, the reliability of the surrogate explanations depends on the extent the surrogate model can reproduce the black-box model's predictions.

Permutation feature importance \cite{b18} is a model-agnostic way to measure feature importance. It works by observing the change in the model's prediction error after permuting the values of a feature. A feature is more important if shuffling its values increases the model error more, implying that the model depends on that feature for making its predictions. A feature is not important if shuffling its importance does not significantly change the model error. 

SHAP \cite{b19} is a game theory-based method that uses Shapley values to explain a model's individual predictions. It works by explaining the prediction of an instance by computing how much each feature contributes to that prediction. A collective overview of the Shapley values enables the global interpretability of the model. Also, the absolute values of the Shapley values can serve as feature importance metric. Then, each feature's effect on the model can be fully explored by examining a SHAP summary plot, which plots individual observations' Shapley values for the most critical features and indicates outcomes with different colors. This broad, global view is still in terms of Shapley's values. To get a deeper look at what these values exactly mean, a SHAP dependence plot can be examined to see a single feature's effect on model predictions. Further, the dependence plot allows for the visualization of feature interactions. SHAP automatically detects which feature has the most significant interaction and colors by that feature in the dependence plot: interactions can be specified for any feature or excluded altogether. Overall, SHAP enables a deeper look into the explanations regarding individual features and predictions, as well as the whole.

ALE \cite{b20} plots depict how a feature influences the model's predicted probabilities as well as the distribution of the feature values. This enables gaining more information about how the feature values affect the models' predictions: essentially a simpler version of SHAP's dependence plot. ALE is similar to the traditional partial dependence plot but is generally faster, unbiased, and less problematic with correlated features \cite{b16, b20, b21}.

\section{Results and Findings}

\subsection{Accuracy}

The predictive performance results obtained on the test set are provided in Table I, with the best values are shown in bold for each metric. Among the eight machine learning models, the best performing model is XGBoost, with an AUC score of $0.81$, an F1 score of $0.73$, and overall accuracy of $0.74$. The best performing inherently interpretable model is LASSO, with an AUC score of $0.77$, an F1 score of $0.70$, and overall accuracy of $0.71$. This difference in performance is notable, although not as stark of a difference as one may expect for the loss in interpretability. 

\begin{table}[htbp]
\caption{Predictive Performance Results}
\begin{center}
\begin{tabular}{|l|c|c|c|}
\hline
\textbf{Model}&\multicolumn{3}{|c|}{\textbf{Metric}} \\
\cline{2-4} 
& \textbf{Accuracy}& \textbf{F1 Score}& \textbf{AUC Score} \\
\hline
Decision Tree&0.69&0.68&0.74\\
\hline
Random Forest&0.73&0.71&0.79\\
\hline
LASSO&0.71&0.70&0.77\\
\hline
Adaptive Boosting&0.72&0.71&0.79\\
\hline
Gradient Boosting&0.73&0.72&0.80\\
\hline
XGBoost&\textbf{0.74}&0\textbf{.73}&\textbf{0.81}\\
\hline
SVM&0.71&0.69&0.77\\
\hline
Neural Network&0.71&0.69&0.76\\
\hline
\end{tabular}
\label{tab1}
\end{center}
\end{table}

\subsection{Fairness}

\begin{table*}[]
\centering
\caption{Bias Audit Results, Gender}
\begin{tabular}{|l|l|l|l|l|l|c|c|} 
\hline
\multicolumn{1}{|c|}{\multirow{2}{*}{\textbf{Model}}} & \multicolumn{6}{c|}{\textbf{Metric}}                                                                                                                                                                                                                                              & \multirow{2}{*}{\begin{tabular}[c]{@{}c@{}}\textbf{\textit{Avg. Dist. }}\\\textbf{\textit{ From Ref.}}\end{tabular}}  \\ 
\cline{2-7}
\multicolumn{1}{|c|}{}                                & \textbf{PPRD}                               & \textbf{PPGRD}                              & \textbf{FDRD}                               & \textbf{FPRD}                               & \textbf{FORD}                               & \textbf{FNRD}                               &                                                                                                                       \\ 
\hline
Decision Tree                                         & \textbf{0.11} & \textbf{0.80} & 1.31                                        & \textbf{0.78} & 0.74                                        & \textbf{1.31} & \textit{\textbf{0.37}}                                                                           \\ 
\hline
LASSO                                                 & 0.08                                        & 0.61                                        & 1.01                                        & 0.46                                        & 0.83                                        & 1.93                                        & \textit{0.49}                                                                                                         \\ 
\hline
Random Forest                                         & 0.10                                        & 0.73                                        & 1.24                                        & 0.67                                        & 0.84                                        & 1.74                                        & \textit{0.44}                                                                                                         \\ 
\hline
Adaptive Boosting                                     & 0.08                                        & 0.59                                        & \textbf{1.00} & 0.44                                        & 0.88                                        & 2.08                                        & \textit{0.52}                                                                                                         \\ 
\hline
Gradient Boosting                                     & 0.09                                        & 0.66                                        & 1.13                                        & 0.55                                        & 0.88                                        & 1.89                                        & \textit{0.47}                                                                                                         \\ 
\hline
XGBoost                                               & 0.09                                        & 0.64                                        & 1.09                                        & 0.53                                        & \textbf{0.89} & 1.98                                        & \textit{0.49}                                                                                                         \\ 
\hline
SVM                                                   & 0.09                                        & 0.63                                        & 1.04                                        & 0.49                                        & 0.83                                        & 1.92                                        & \textit{0.49}                                                                                                         \\ 
\hline
Neural Network                                        & 0.09                                        & 0.61                                        & 0.98                                        & 0.45                                        & 0.81                                        & 1.93                                        & \textit{0.50}                                                                                                         \\ 
\hline
\textit{Average}                                      & \textit{0.09}                               & \textit{0.66}                               & \textit{1.10}                               & \textit{0.55}                               & \textit{0.84}                               & \textit{1.85}                               &                                                                                                                       \\
\hline
\end{tabular}
\end{table*}

\begin{table*}[]
\centering
\caption{Bias Audit Results, Race}
\begin{tabular}{|l|l|l|l|l|l|c|c|} 
\hline
\multicolumn{1}{|c|}{\multirow{2}{*}{\textbf{Model}}} & \multicolumn{6}{c|}{\textbf{Metric}}                                                                                                                                                                                                                                              & \multirow{2}{*}{\begin{tabular}[c]{@{}c@{}}\textit{\textbf{Avg. Dist. }}\\\textit{\textbf{ From Ref.}}\end{tabular}}  \\ 
\cline{2-7}
\multicolumn{1}{|c|}{}                                & \textbf{PPRD}                               & \textbf{PPGRD}                              & \textbf{FDRP}                               & \textbf{FPRD}                               & \textbf{FORD}                               & \textbf{FNRD}                               &                                                                                                                       \\ 
\hline
Decision Tree                                         & 1.58                                        & 1.14                                        & 1.06                                        & 1.26                                        & \textbf{0.99} & 0.78                                        & \textit{0.21}                                                                                                         \\ 
\hline
LASSO                                                 & \textbf{1.52} & \textbf{1.10} & \textbf{1.05} & \textbf{1.20} & 1.02                                        & 0.84                                        & \textit{\textbf{0.18}}                                                                  \\ 
\hline
Random Forest                                         & \textbf{1.52} & \textbf{1.10} & \textbf{1.05} & \textbf{1.20} & \textbf{0.99} & 0.80                                        & \textit{\textbf{0.18}}                                                                  \\ 
\hline
Adaptive Boosting                                     & 1.57                                        & 1.13                                        & 1.08                                        & 1.27                                        & 0.98                                        & 0.77                                        & \textit{0.22}                                                                                                         \\ 
\hline
Gradient Boosting                                     & \textbf{1.52} & \textbf{1.10} & 1.08                                        & 1.24                                        & 1.04                                        & \textbf{0.86} & \textit{0.19}                                                                                                         \\ 
\hline
XGBoost                                               & 1.55                                        & 1.12                                        & 1.11                                        & 1.30                                        & \textbf{1.01} & 0.81                                        & \textit{0.21}                                                                                                         \\ 
\hline
SVM                                                   & \textbf{1.52} & \textbf{1.10} & 1.06                                        & 1.21                                        & 1.03                                        & 0.85                                        & \textit{\textbf{0.18}}                                                                  \\ 
\hline
Neural Network                                        & 1.53                                        & 1.11                                        & 1.08                                        & 1.25                                        & 1.05                                        & 0.84                                        & \textit{0.20}                                                                                                         \\ 
\hline
\textit{Average}                                      & \textit{1.54}                               & \textit{1.11}                               & \textit{1.07}                               & \textit{1.24}                               & \textit{1.01}                               & \textit{0.82}                               &                                                                                                                       \\
\hline
\end{tabular}
\end{table*}

\subsubsection{Introduction}

In the NIJ recidivism data, two protected classes must be examined for fairness: gender and race. The NIJ data provides binary variables for both: male or female for gender and White or Black for the race. Unfortunately, the dataset is already imbalanced concerning gender and race, having more males than females (88\% male, 12\% female) and more Black individuals than White individuals (57\% Black, 43\% White). As a result, the data may already have inherent issues before models are trained on it.

The results of the bias audits are provided in Table II for gender and Table III for race, with males and White as the reference groups, respectively. The best score for each test is denoted in bold. Each score represents relative position of the model with respect to the parity for the corresponding test. A score of $1$ means perfect parity, but it is standard to have an acceptable fairness interval from $0.80$ to $1.25$ \cite{b22}. While reading the scores, it should be remembered that the reference groups are male and White. Then, each metric score can be read as a percentage; for example, a score of $1.2$ for \emph{False Positive Rate Disparity} for race indicates that the False Positive Rate for Black individuals is 120\% of the value of the False Positive Rate for White individuals (reference group), which falls within the acceptable fairness range. The final column in the tables, \emph{Average Distance From Reference}, is the mean of the entire row for each model, representing the average disparity across the different bias tests. This metric can find the ``overall'' least biased model. However, it does not capture the entire picture because some tests have more egregious bias than others, depending on the objective of the ML task at hand.

\subsubsection{Gender}

In terms of gender unbiasedness, the best-performing model is the decision tree, as it has the closest value to parity for four out of six tests and has the lowest overall distance to the reference. Although it is the clear winner, only two of the scores fall within the acceptable fairness range, meaning that even the best model with respect to gender fairness has serious problems.

The decision tree is also the worst performing model overall regarding the accuracy and the most interpretable, leading to questions about the trade-off between accuracy, fairness, and interpretability. The best overall performing model, XGBoost, has an average distance of $0.49$ from the reference, which is the same average value for the interpretable LASSO and the worse performing but still less interpretable SVM, indicating that the trade-off is not clear-cut. The rest of the models have an average distance from reference significantly larger than the decision tree, but all around the same amount. 

Overall, if a model needs to be selected that optimized gender fairness, interpretability, and accuracy, a good choice would be the random forest. The random forest is comparatively simpler to interpret than most black-box methods as it makes its predictions based on a simple majority vote of independent decision trees rather than the more complicated methods that other black-box models use. Further, the random forest has close to the best accuracy and is the second most fair model overall with respect to gender. 

\subsubsection{Race}

Unlike gender unbiasedness, there is no clear winner for the race, as three models have an average distance of $0.18$ from the reference: the random forest, LASSO, and SVM. Each of the models has a different distribution across the six test metrics, which makes the ``winner'' for bias concerning race dependent on whatever metric is deemed the most relevant for the task at hand.

SVM would likely be eliminated, if a model must be selected, due to the comparatively lower accuracy and the more complex interpretation. The decision between the random forest and LASSO would then depend on whether accuracy or interpretability matters more to the user. If accuracy matters more, select the random forest. If interpretability matters more, select LASSO.

\subsubsection{Overview}

Overall, the disparities for the race are smaller than the disparities for gender: perhaps because there is more of a gender imbalance in the data than a race imbalance. The disparities are also different among models for both protected classes; if one ordered the models based on average distance from reference scores, the only model in the same order for gender and race would be adaptive boosting with the worst overall distance. 

If a model needs to be selected to account for gender and race bias, the random forest would be a good choice. It is tied for the best model concerning race fairness and the clear runner-up concerning gender fairness. Further, the random forest model still has a good accuracy and while not inherently explainable, its interpretations are simpler than boosting-based tree models or neural networks.

\subsection{Interpretability}

\subsubsection{Inherently Interpretable}

Unfortunately, the decision tree fitted was too large to fit onto this paper as it has 64 leaf nodes. Instead, Gini importances were analyzed. In the decision tree fitted, the most important features are: 

\renewcommand{\theenumi}{\alph{enumi}}
\begin{enumerate}
  \item \emph{Percent\_Days\_Employed}, the percentage of days an individual was employed while on parole.
  \item \emph{Jobs\_Per\_Year}, the number of different jobs held while on parole.
  \item \emph{Prior\_Arrest\_Episodes\_PPViolationCharges}, whether an individual was previously arrested for violating their parole or probation conditions.
  \item \emph{Gang\_Affiliated\_True}, if an investigation verified a gang affiliation.
\end{enumerate}

These four features account for 76.4\% of the overall feature importance in the decision tree, meaning they are the most meaningful out of the 62 features used. Further, the top 11 features (provided in Figure 1) account for more than 99\% of the total feature importance. Therefore, a decision tree with fewer variables, rather than the entire 62, would be even easier to interpret and perform similarly. Around 30 features have importances smaller than $0.0001$ meaning that they provided virtually no additional information to the decision tree's learning. 

\begin{figure}[htbp]
\centerline{\includegraphics[width=0.475\textwidth]{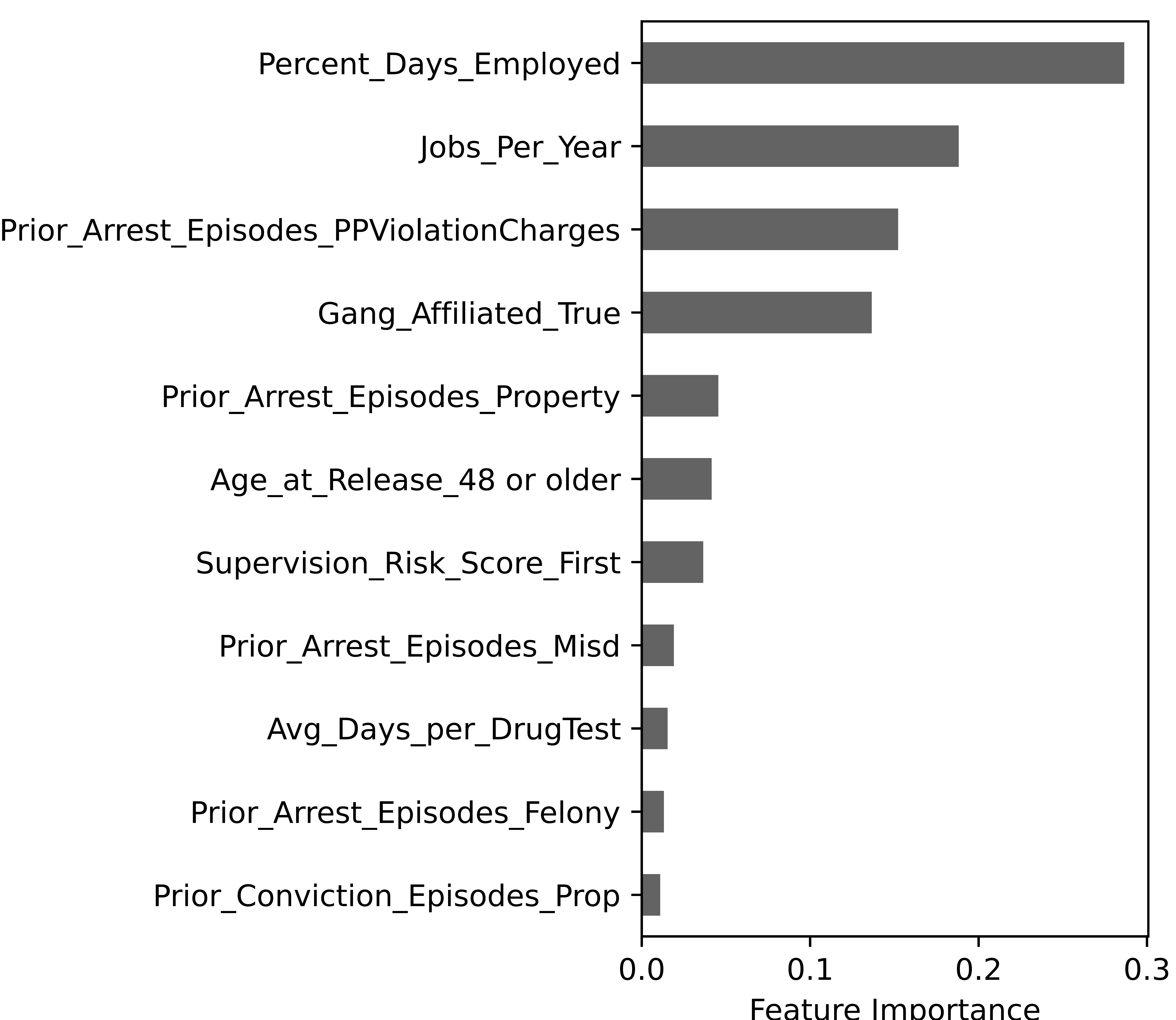}}
\caption{Gini feature importances for the most critical 11 features in the decision tree.}
\label{fig}
\end{figure}

Unfortunately, decision tree feature importances cannot reveal how much each feature contributed to each outcome. If having a model where this is mathematically clear is desired, LASSO can be used. The 11 largest coefficients in the LASSO model (to loosely compare to the 11 most important features in the decision tree) are provided in Table IV. Although most features' value ranges are similar, they are neither the same nor standardized. Thus, while LASSO coefficients are helpful, they do not directly correspond to the feature importances.

\begin{table}[htbp]
\caption{LASSO Model Coefficients (Top 11)}
\begin{center}
\begin{tabular}{|l|c|}
\hline
\textbf{Feature}&{\textbf{Coefficient}}\\
\hline
\emph{DrugTests\_Meth\_Positive}&\phantom-2.12\\
\hline
\emph{Percent\_Days\_Employed}&-1.57\\
\hline
\emph{Age\_at\_Release\_48\_or\_older}&-1.54\\
\hline
\emph{Age\_at\_Release\_43-47}&-1.22\\
\hline
\emph{Age\_at\_Release\_38-42}&-1.04\\
\hline
\emph{Age\_at\_Release\_33-37}&-0.93\\
\hline
\emph{DrugTests\_THC\_Positive}&\phantom-0.78\\
\hline
\emph{Gang\_Affiliated\_True}&\phantom-0.76\\
\hline
\emph{Age\_at\_Release\_28-32}&-0.63\\
\hline
\emph{DrugTests\_Cocaine\_Positive}&\phantom-0.51\\
\hline
\emph{Jobs\_Per\_Year}&\phantom-0.47\\
\hline
\end{tabular}
\label{tab1}
\end{center}
\end{table}

LASSO makes its decisions by creating a threshold boundary at the probability of $0.5$, the midpoint between $0$ (False) and $1$ (True). If the value for an individual's probability is less than $0.5$, they will be classified as False, meaning that the model predicts that they will not commit a crime. The opposite is true for values greater than or equal to $0.5$. The coefficients ($c$) come into play by changing the estimated odds for an individual by a factor of $e^{c}$. Thus, for example, a one unit increase in \emph{DrugTests\_Meth\_Positive} results in an increase in the estimated odds for the individual by $e^{2.12}$. Therefore, it can be observed in what direction and by how much each feature affects the final prediction. 

It is interesting to compare the features in the two interpretable models. Among the 11 most important features/largest coefficients, the two models shared \emph{Percent\_Days\_Employed}, \emph{Age\_at\_Release\_48\_or\_older}, \emph{Gang\_Affiliated\_True}, and \emph{Jobs\_Per\_Year}. LASSO learned more from drug-related features as LASSO emphasized \emph{DrugTests\_Meth\_Positive}, \emph{DrugTests\_THC\_Positive}, and \emph{DrugTests\_Cocaine\_Positive} which all indicate the proportion of drug tests taken that were positive for a certain drug. The decision tree gave almost no importance to all drug-related features. Similarly, LASSO assigned larger coefficients to several age-related features, while the decision tree only found considerable importance in the oldest age-related feature. The decision tree also found more importance in crime-related features such as \emph{Prior\_Arrest\_Episodes\_PPViolationCharges} and \emph{Prior\_Arrest\_Episodes\_Property}, which LASSO did not assign high coefficient values for. 

Based on the above analysis, it appears that the two inherently interpretable models rely on a different set of features from each other in making their predictions. As can be seen in Table I, LASSO has better performance metrics than the decision tree, but not by a very large margin. The predictive performances of the not inherently interpretable models are not much better, although they are a noticeable improvement over the inherently interpretable models. Is this slight improvement worth the loss in interpretability?

\subsubsection{Not Inherently Interpretable}

In this subsection, XGBoost, the overall best performing model, will be interpreted using various interpretability methods. XGBoost is a tree-based boosting method, meaning it uses many small trees that build off each other to make a decision. Therefore, general interpretability methods fail.

First, a LASSO model is trained to serve as a global surrogate model and is examined. The $R^2$ value between the XGBoost predictions and the surrogate model predictions on the test set is $0.38$. The surrogate model only explains $38\%$ of the variance in the XGBoost model's predictions. Therefore, the surrogate model is considered a poor explainer for XGBoost. However, the surrogate model is still interpreted to see what it suggests. The five most significant coefficients for the surrogate model are provided in Table V.

\begin{table}[htbp]
\caption{Model Coefficients for the Surrogate LASSO Model (Top 5)}
\begin{center}
\begin{tabular}{|l|c|}
\hline
\textbf{Feature}&{\textbf{Coefficient}}\\
\hline
\emph{DrugTests\_Meth\_Positive}& \phantom-5.82\\
\hline
\emph{Percent\_Days\_Employed}&-4.02\\
\hline
\emph{Age\_at\_Release\_48\_or\_older}&-2.63\\
\hline
\emph{Gang\_Affiliated\_True}&\phantom-1.96\\
\hline
\emph{Age\_at\_Release\_43-47}&-1.88\\
\hline
\end{tabular}
\label{tab1}
\end{center}
\end{table}

In the surrogate LASSO model, each feature also had a high coefficient value in the original LASSO model, although not in the same order and with different magnitudes. The interpretation is the same as the original LASSO model; for example, a one unit increase in \emph{DrugTests\_Meth\_Positive} increases the estimated odds of the individual committing a crime while on parole by $e^{5.82}$. Therefore, the surrogate model suggests that XGBoost may rely on many of the same features as the LASSO model, although with a higher emphasis on a few. However, this interpretation should be taken with caution since the $R^2$ value for the surrogate model is low. 

Next, permutation feature importances are examined. The five most important found features are shown in Figure 2. Both \emph{Percent\_Days\_Employed} and \emph{Gang\_Affiliated\_True} also appeared in the five highest coefficients in the surrogate model. \emph{Avg\_Days\_Per\_DrugTest} and \emph{Delinquency\_Reports} are two features that have not shown up in top features for importances or significant coefficients up to this point. The former feature describes how often a parolee is tested for drugs, and the latter describes how many parole delinquency reports (for minor violations) have been received for a parolee. 

\begin{figure}[htbp]
\centerline{\includegraphics[width=0.5\textwidth]{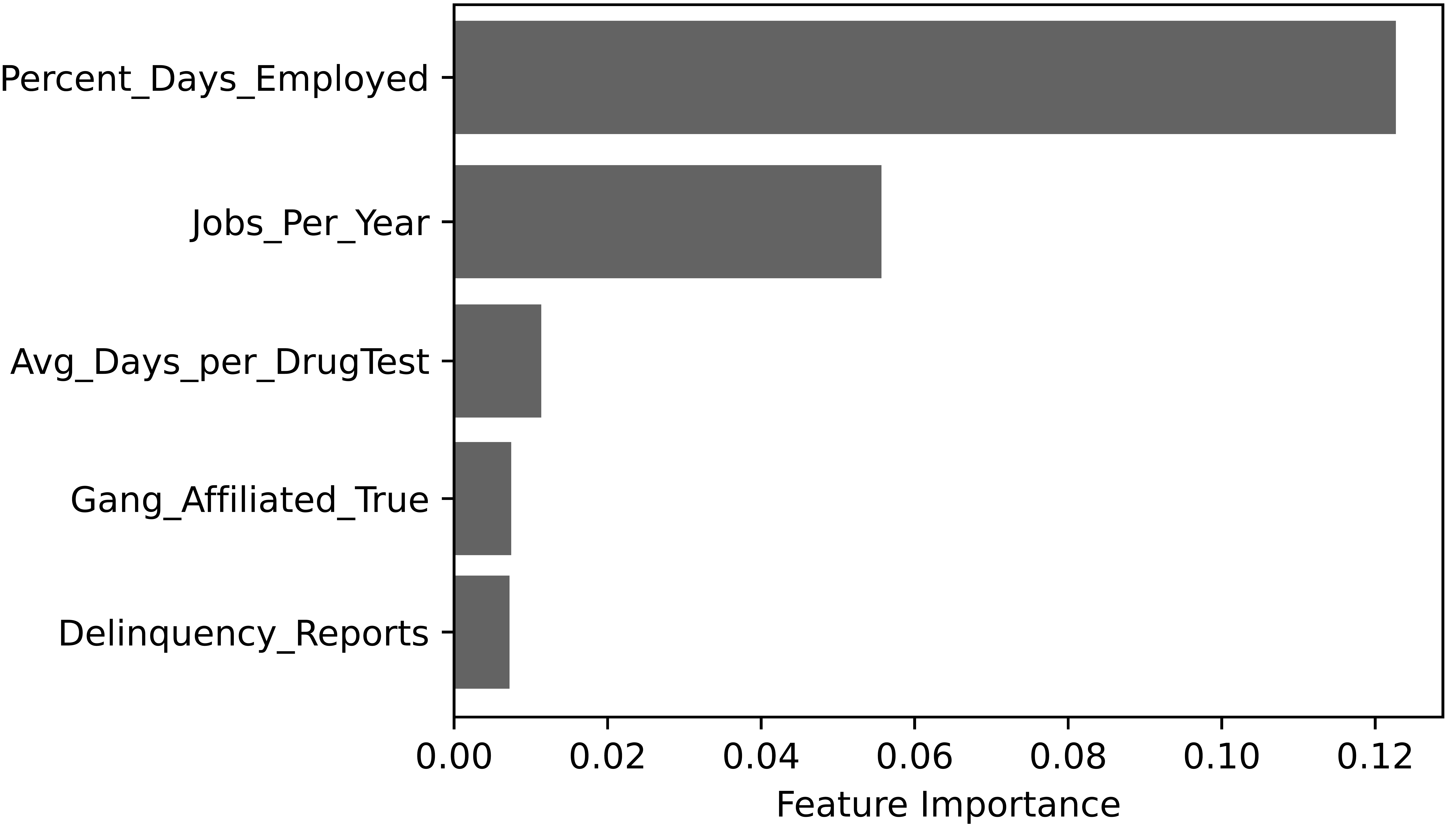}}
\caption{The five largest permutation feature importances for the best performing model, XGBoost.}
\label{fig}
\end{figure}

An interpretability analysis is also conducted via SHAP. The five most important features based on absolute SHAP values are provided in Figure 3. The corresponding summary plot is provided in Figure 4. Each point is colored for outcome: fuchsia for predicted recidivism and blue otherwise. The points for a specific feature are jittered vertically for visualization purposes. Any overlap of points with different colors indicates the lack of a sharp distinction for that range of Shapley values. A clear pattern can be seen for most features: Shapley values on either side of the zero point are correlated with a specific outcome. For example, positive Shapley values for \emph{Jobs\_Per\_Year} are correlated with recidivism.

The SHAP dependence plot for \emph{Jobs\_Per\_Year} is given in Figure 5. A non-linear but generally monotonically increasing pattern can be observed. This implies that the larger the value of \emph{Jobs\_Per\_Year}, the higher its Shapley value in general. Then, it is possible to go back to the summary plot in Figure 4 and observe that the negative Shapley values correspond to having relatively fewer jobs in a year. This is, then, correlated with not committing a crime again. This is perhaps an ``expected'' pattern for the \emph{Jobs\_Per\_Year} feature: having a large number of jobs per year could indicate instability in a parolee's life which common sense suggests would be a risk for recidivism. Figure 5 also provides feature interactions for \emph{Percent\_Days\_Employed}. The points are colored by the value of \emph{Percent\_Days\_Employed} across two bins: values less than $0.5$ and greater than or equal to $0.5$. In the figure, it can be observed that for the lower range of \emph{Jobs\_Per\_Year}, \emph{Percent\_Days\_Employed} values of less than $0.5$ increases the risk of recidivism as blue-colored points are above the fuchsia-colored points, resulting in larger SHAP values for blue-colored points for the same values of \emph{Jobs\_Per\_Year}.

\begin{figure}[htbp]
\centerline{\includegraphics[width=0.47\textwidth]{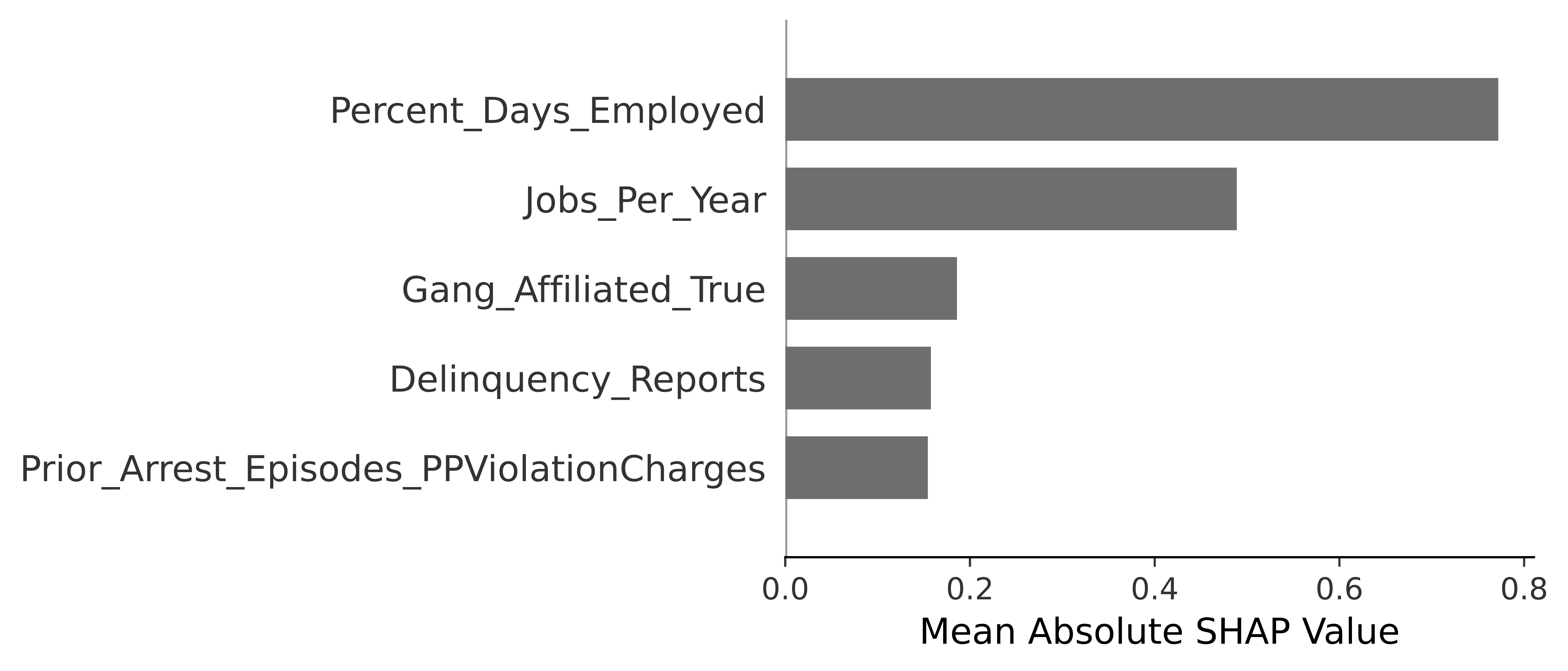}}
\caption{The five largest SHAP feature importances for the XGBoost model.}
\label{fig}
\end{figure}

\begin{figure}[htbp]
\centerline{\includegraphics[width=0.47\textwidth]{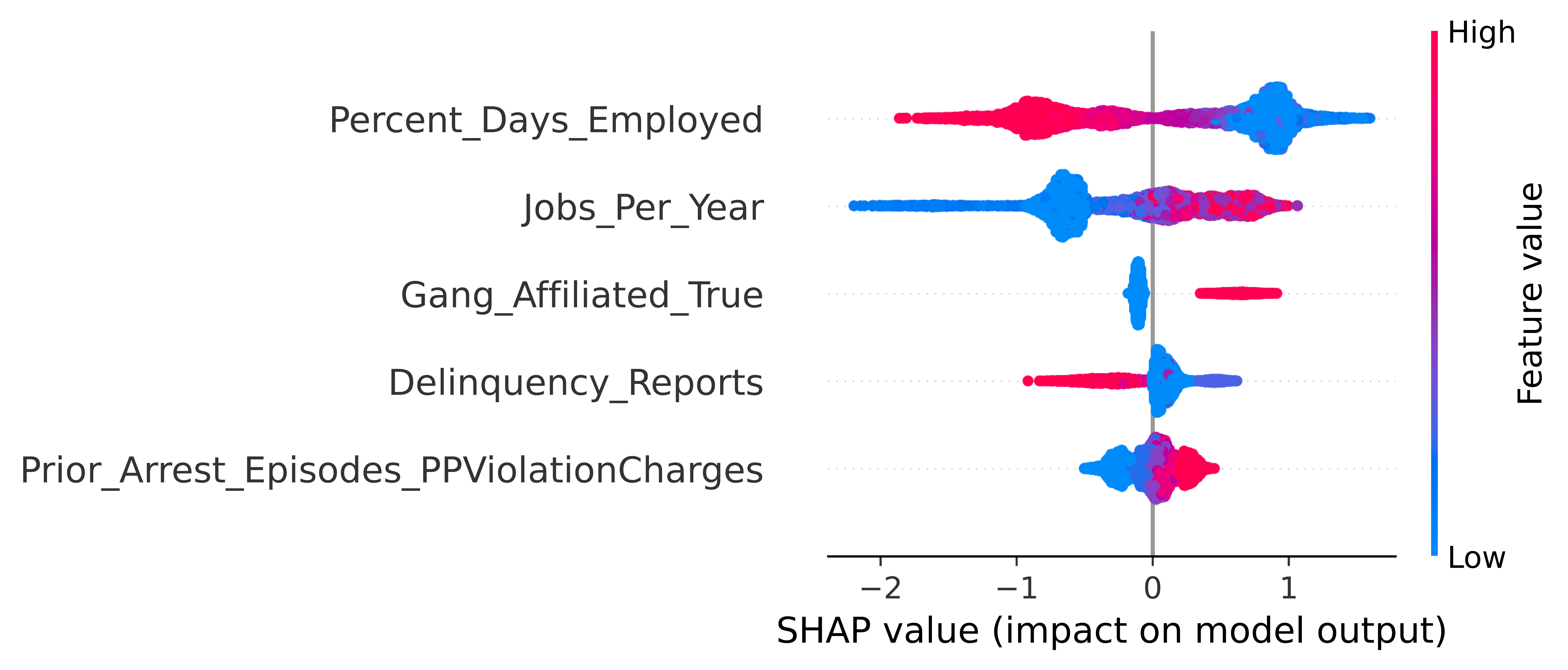}}
\caption{The SHAP summary plot for the five most important features (based on SHAP feature importances) for the XGBoost model.}
\label{fig}
\end{figure}

\begin{figure}[htbp]
\centerline{\includegraphics[width=0.44\textwidth]{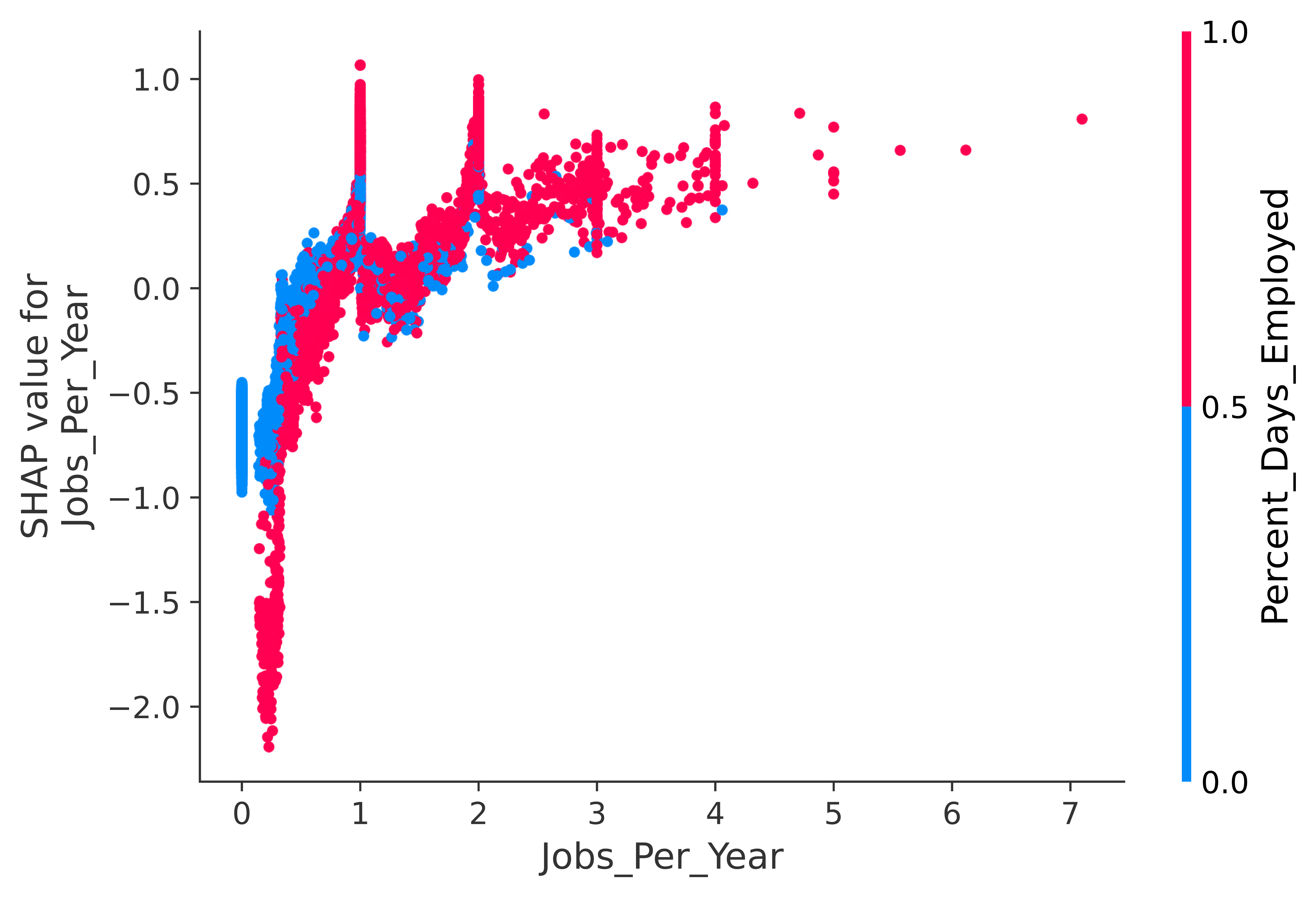}}
\caption{The SHAP dependence plot for \emph{Jobs\_Per\_Year}. The points are colored based on the values for \emph{Percent\_Days\_Employed}.}
\label{fig}
\end{figure}

Finally, Accumulated Local Effect (ALE) plots are examined. Throughout the analyses on the feature importances and coefficients seen with the surrogate model, SHAP, and permutation feature importance, certain features are more frequently identified as contributing the most to the model predictions. ALE plots can be created for those features.

Figure 6 shows the ALE plot for \emph{Percent\_Days\_Employed}. The bottom of the plot is shaded to show the distribution of the feature: in this case, there is a fairly even distribution. As a result, the confidence interval on the plot is pretty narrow. The plot shows that, for its values from around $0$ to $0.2$, the feature strongly affects the prediction probability of recidivism. Then, from $0.2$ to $1$, the effect on the prediction probability shifts towards an individual not committing a crime on parole in a roughly negative, quadratic trend.

Figure 7 shows the ALE plot for \emph{DrugTests\_Meth\_Positive}. Unlike the ALE plot for \emph{Percent\_Days\_Employed} in Figure 6, the data is not evenly distributed, so the confidence interval is relatively wide, but the feature still has a noticeable effect on the prediction. The plot is also largely positive, showing that having positive tests for meth only increases the chance that the prediction will be that the parolee commits a crime. Further, the relationship has two different linear trends: a steeper linear trend for where the data is concentrated and a less steep linear trend for where the data is scattered.

\begin{figure}[htbp]
\centerline{\includegraphics[width=0.48\textwidth]{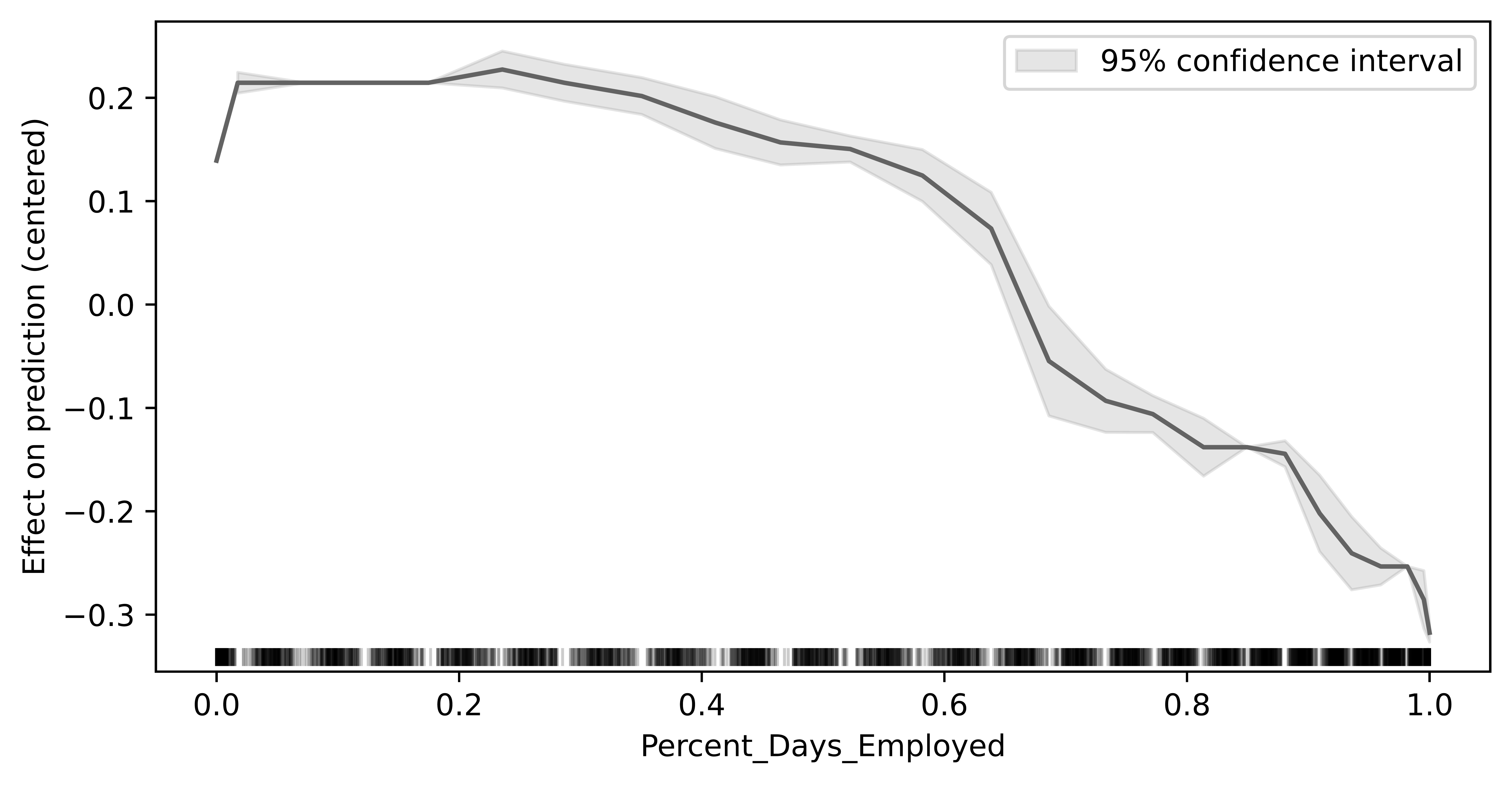}}
\caption{The ALE plot for the variable \emph{Percent\_Days\_Employed}.}
\label{fig}
\end{figure}

\begin{figure}[htbp]
\centerline{\includegraphics[width=0.48\textwidth]{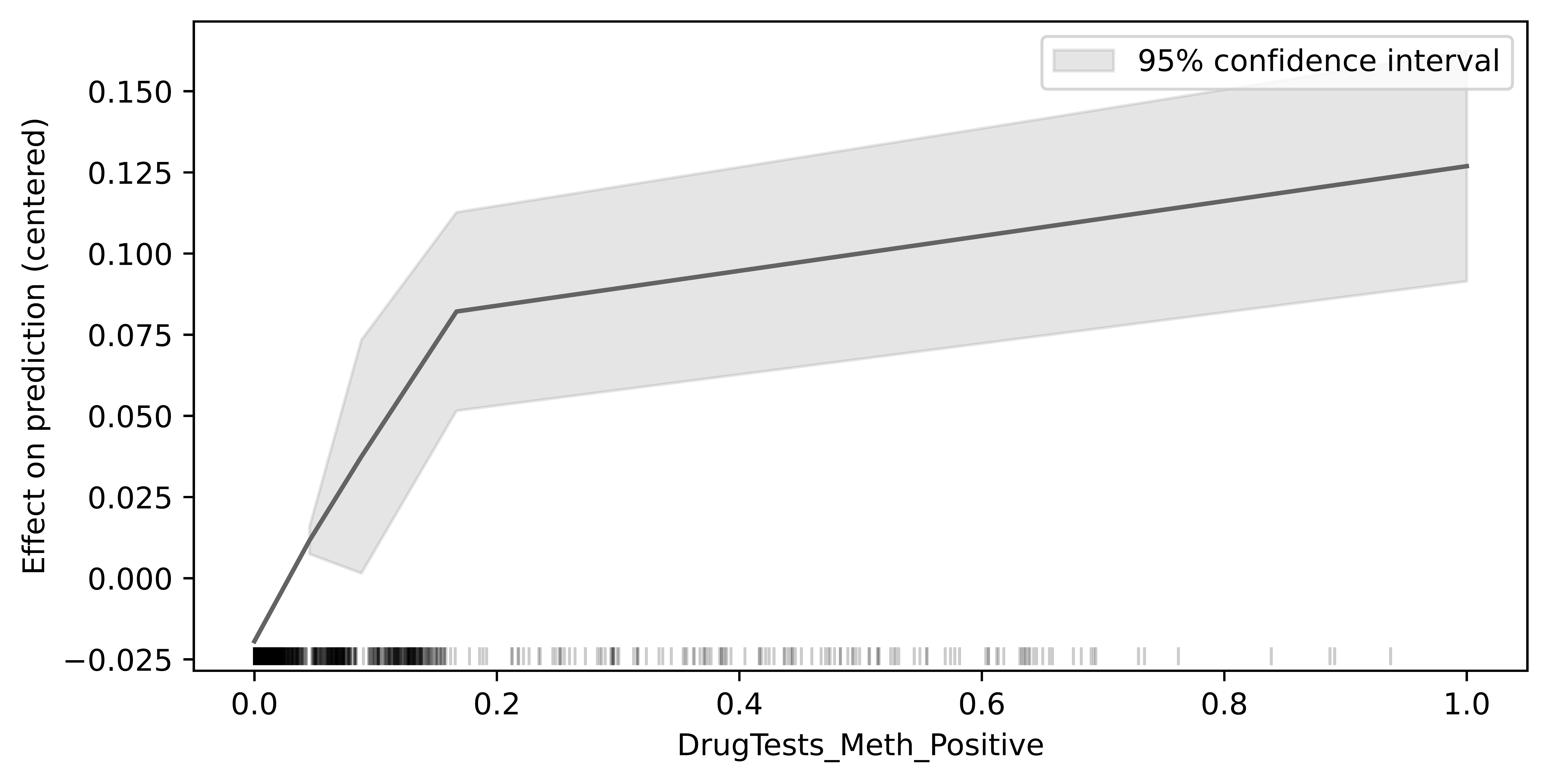}}
\caption{The ALE plot for the variable \emph{DrugTests\_Meth\_Positive}.}
\label{fig}
\end{figure}

\section{Conclusion}

Criminal recidivism models are used as a factor in parole boards' decision-making in parole hearings; however, these models still have many unanswered questions about their accuracy, fairness, and interpretability. This paper created several recidivism models using real-world data from the state of Georgia. Subsequently, it examined their accuracy, fairness, and interpretability to provide  insights into the characteristics of recidivism models beyond predictive performance. 

This work has some limitations. First, missing data imputation methods may have impacted the results, particularly the interpretability analysis. Second, the generalizability of the findings depends on similar studies conducted on other datasets. Third, a detailed exploration into features that are found to be less critical merits consideration. Removing these features would improve the ease of interpretability but may also affect accuracy and fairness. Given the importance of such critical systems, future work may tackle these limitations and improve real-world decision-aiding systems in criminal justice.

There are noted differences between the machine learning models employed in this paper regarding accuracy, fairness, and interpretability. No model is found to be consistently the best in all three aspects, but several general trends are identified. Regarding accuracy, tree-based boosting methods (which are not inherently interpretable and are considered black-box) performed very well, with the overall best being XGBoost. However, LASSO, an inherently interpretable model, performed better than two black-box methods, SVM and neural network. These findings indicate that the accuracy-interpretability trade-off is not always clear-cut.

Moreover, black-box models showed many of the same tendencies as the inherently interpretable models in terms of feature importance and coefficients, a focus on features related to employment, drug use, prior arrests, gang affiliation, and age. In regards to fairness, none of the models would meet the standard criteria for fairness. This is possibly due to the imbalance in the data: gender, which was more imbalanced than race, fell further outside the standard metric. Regardless, the most ``fair'' model for gender was the decision tree, followed by the random forest. The most ``fair'' models for the race were the random forest, SVM, or LASSO. The more accurate and less interpretable models tended to perform worse for fairness: no boosting model (i.e., the most accurate models) performed comparatively well in fairness. Therefore, it can be concluded that all models have shortcomings, and the suitable criminal recidivism model for the policymakers in Georgia to choose is a question of desired balance between accuracy, fairness, and interpretability.

\section*{Acknowledgment}
The authors thank Dr. Ryan Kennedy for his valuable contributions to this study.
The first author's work was supported by the University of Houston's Computer Science REU program that is primarily sponsored by the National Science Foundation under Award CCF-195029 and the University of Houston's College of Natural Sciences and Mathematics. Drs. Gursoy and Kakadiaris’ work was supported by the National Science Foundation under Award CCF-2131504. Any opinions, findings, and conclusions or recommendations expressed in this material are those of the author(s) and do not necessarily reflect the views of the National Science Foundation.

\end{document}